\documentstyle[12pt]{article}

\def\PL #1 #2 #3 {{\it Phys. Lett.} {\bf#1} (#3) #2}
\def\NP #1 #2 #3 {{\it Nucl. Phys.} {\bf#1} (#3) #2}
\def\ZP #1 #2 #3 {{\it Z. Phys.} {\bf#1} (#3) #2}
\def\PRL #1 #2 #3 {{\it Phys. Rev. Lett.} {\bf #1} (#3) #2}
\def\PR #1 #2 #3 {{\it Phys. Rev.} {\bf#1} (#3) #2}
\def\MPL #1 #2 #3 {{\it Mod. Phys. Lett.} {\bf#1} (#3) #2}
\def\RMP #1 #2 #3 {{\it Rev.~Mod. Phys.} {\bf#1} (#3) #2}

\begin{document}
\setlength{\baselineskip}{2.4ex}
\begin{flushright}
Fermilab-Conf-94/260-T\\
FSU--HEP--940817\\
August 1994
\end{flushright}

\begin{center}
{\bf $W$ PLUS HEAVY QUARK PRODUCTION AT THE TEVATRON
\footnote[2]{Presented at DPF Meeting, Albuquerque, New Mexico,
August 1-6,1994}}\\
S.~KELLER\footnote[3]{presenting author}\\
{\em Physics Department, B-159\\ Florida State University\\
Tallahassee, FL 32306}\\
\vspace{0.3cm}
and\\
\vspace*{0.3cm}
W.~T.~GIELE and E.~LAENEN\\
{\em Fermi National Accelerator Laboratory\\ P.O. Box 500\\
Batavia, IL 60510, USA}\\
\end{center}

\begin{center}
\parbox{13.0cm}
{\begin{center} ABSTRACT \end{center}

{\small We summarize the motivations for and the status of the
 calculation of the $W +$ heavy quark production process in $p \bar p$
colliders to
Next-to-Leading Order in QCD.  This process can be used to constrain the
strange quark distribution function at high $Q^2$ at the Tevatron, and also to
study the bottom content of $W+1$~jet events.  In addition, when crossed, the
calculation essentially describes the single top quark production process to
Next-to-Leading Order in QCD.}

}

\end{center}

\section{Introduction}

There are well-known benefits to a Next-to-Leading Order (NLO) calculation
over a Leading Order (LO) one: the dependence on the
renormalization and factorization scales is
reduced; the parton shower starts to be reconstructed; and
the calculation begins to be sensitive to detector limitations.
Furthermore, the NLO calculation checks the validity of
the LO one, and thus the validity of the perturbative expansion.

The motivations for the NLO calculation
of the $W$ plus heavy quark production process
and its status
are summarized in  Section 2 and 3, respectively.

\section{Motivations}

\subsection{$W + charm$}

This is a summary of the analysis done in Ref.~\ref{BHKMR} with the shower
Monte-Carlo program Pythia~\cite{PYT87}.
At low $Q^2$ the strange quark distribution function can be measured with
the appropriate linear combination of $F_2$ structure functions in neutrino
and muon deep inelastic scattering~\cite{CTE93}.
It can also be measured from di-muon
events in neutrino deep-inelastic scattering~\cite{CCF92}.
Using current
experimental data sets, the two methods yield
a difference of about a factor of 2
for the strange quark distribution.
In Ref.~\ref{BHKMR} it was suggested that the strange quark distribution
function
can also be constrained by determining the charm content of $W+1$~jet events,
because the underlying
subprocess, $sg \rightarrow W+c$, is directly proportional to
the strange quark distribution function.  In this measurement the strange quark
will be probed at large $Q^2 \simeq M_W^2$, and will therefore provide a
consistency
check with lower $Q^2$ measurements.
At $Q^2 = M_{W}^2$ the difference between the two different strange quark
distribution functions is smaller due to evolution.
The bottom line is that when
the relevant backgrounds are included and standard cuts are used,
the factor of 2 becomes a difference of about 14\%.  The tagging
efficiency needed for the statistical uncertainty to
equal this is around 10 \% for 6000 $W+1$~jet events.
There are three tagging methods available:  reconstruction
of a secondary vertex using an SVX~\cite{SVX90},
direct reconstruction of the decayed D-meson, and the
tagging of a lepton in the semi-leptonic decay\cite{RH}
of the D-meson.  When combined,
these methods are likely to achieve the required tagging efficiency.
We refer the reader to Ref.~\ref{BHKMR} for further details.

\subsection{$W + bottom$}

As is well known, the $W+n$~jet process with $b$ tag is a background for
the top quark
analysis.  Extensive studies of this background have been done at LO
and with shower Monte-Carlo programs.
Clearly, a comparison of the data with a
NLO calculation in the case $n=1$ will be important.
Furthermore, at large $P_T$, it will be necessary to include
fragmentation functions.

\subsection{$W + top$}

The subprocess $b+g \rightarrow W+t$ is a small contribution to the single
top production. However, if the $W$ is crossed to the initial state and the
$b$ to the final state and a quark leg is added to the $W$,
one of the main contributions (esp. at LHC or larger energies)
to single top production is obtained: $q+g\rightarrow q'+t+\bar b$.
Our calculation contains all ingredients for the NLO
analysis of the $W+g \rightarrow t +\bar b$ process.
At NLO the corrections to the $W$-current quark vertex decouple from the rest
of
the graph, see Ref.~\ref{HAN92}.

\section{Status of the NLO calculation}

\subsection{Virtual corrections}

These consist of the interference between the Born diagrams
and the one-loop self-energy corrections,
vertex corrections, and box diagrams.
Some of the contributions have singularities, and in order to regularize
them, we calculated all graphs in $d$ dimensions.
A $d$-dimensional
Passarino-Veltman reduction formalism was used\footnote{We used
the algebraic manipulation program FORM (Ref.~\ref{FORM})
for much of the algebra.}
to reduce tensor- and
vector-like integrals into scalar integrals.
We calculated various 3- and 4-point scalar functions that were not available
in the literature (due to the presence of the heavy quark and $W$ masses).
The ultraviolet singularities were absorbed through coupling-constant and mass
renormalization.   The part of the expression containing soft and collinear
singularities factorized into a universal factor multiplying the Born cross
section, because there is only one color flow for
the Born diagram; in case there are several color flows, there is a different
factor for each ordered subamplitude, see Ref.~\ref{GG92}.

\subsection{Real corrections}

The contributions from all the 2 to 3 processes have to be included.
Some of them exhibit soft and collinear singularities.
There are several equivalent methods to deal with these
singularities, see Ref.~\ref{WAL93}.
These methods basically consist of separating the
multi-parton phase space into a hard
region, containing no singularities, and a
region in which at least one
of the partons is soft or emitted collinearly.
In the hard phase space region, one
can work in 4 dimensions and perform the integration numerically.
In the soft region, the integration is done analytically,
and the result again factorizes into a universal K-factor multiplying
the Born cross
section.  In the present case, we followed the methods of
Ref. ~\ref{GG92}
and derived the dependence of the K-factor on the mass of the quark.
The separation of the phase space depends on one
parameter (sometimes two).
Each of the contributions will depend on this unphysical
parameter. Of course any observable should not.
The initial state collinear singularities are factorized into the
distribution functions.  The remaining singularities
cancel against the leftover singularities of the virtual corrections.

Once the virtual and real corrections are summed,
the cross section is constituted from terms proportional to the
Born cross section, the finite part of the virtual cross section and the hard
phase space part of the real corrections.
At the present time, we have calculated all of the above contributions
to $W$ plus heavy quark production, and are in the process of
constructing the Monte-Carlo program. It will not include
detailed information on the decay of the $W$ boson.

This research was
supported in part by the  Texas National
Research Laboratory Commission under grant no. RGFY9273, and by the
U.S. Department of Energy under
contract number DE-FG05-87ER40319.


\begin{thebibliography}{9}

\bibitem{BHKMR}\label{BHKMR}
U.~Baur, F.~Halzen, S.~Keller, M.~L.~Mangano, and K.~Riesselmann,
\PL  B318 544 1993 .

\bibitem{PYT87}\label{PYT87}
H.--U.~Bengtsson and T.~Sj\"ostrand, {\it Computer Physics Commun.} {\bf
46}
(1987) 43, \\
T.~Sj\"ostrand, CERN-TH.6488/92 preprint.


\bibitem{CTE93}\label{CTE93}
J.~Botts {\it et al.} (CTEQ Collaboration), \PL 304B 159 1993 .

\bibitem{CCF92}\label{CCF92}
W.~H.~Smith {\it et al.} (CCFR Collaboration), preprint WISC--EX--92--326,
to be published in {\it Nucl. Phys.} {\bf B} (Proc. Suppl.).

\bibitem{SVX90}
F.~Abe {\it et al.} (CDF Collaboration), {\it Nucl. Instrum. Methods}
{\bf A289} (1990) 388; \NP B27 246 1992 \ (Proc. Suppl);

\bibitem{RH}\label{RH}
R.~Hamilton, CDF Collaboration, these proceedings, FERMILAB-CONF-94/209-E.

\bibitem{HAN92}\label{HAN92}
T.~Han, G.~Valencia, S.~Willenbrock, \PRL 69 3274 1992 .

\bibitem{FORM}\label{FORM}
FORM2 by J.A.M. Vermaseren, Published by
Computer Algebra Netherlands (CAN), Kruislaan 413,
1098 SJ Amsterdam, The Netherlands.

\bibitem{GG92}\label{GG92}
W.~T.~Giele and E.~W.~N.~Glover, \PR D46 1980 1992 .

\bibitem{WAL93}\label{WAL93}
W.~T.~Giele, E.~W.~N.~Glover, and D.~A.~Kosower, \NP B403 633 1993 \  and
references
therein.  H.~Baer, J.~Ohnemus, and J.~F.~Owens, \PR D40 2844 1989 .



\end{thebibliography}
\end{document}